\documentclass[%
 aip,
 jmp,%
 amsmath,amssymb,
 reprint,%
]{revtex4-1}

\usepackage{graphicx}
\usepackage{dcolumn}
\usepackage{bm}

\begin{document}

\preprint{AIP/123-QED}

\title{Quantum Algorithms for Unit Group and principal ideal problem}

\author{Hong Wang}
\author{Zhi Ma}%
 \email{fallmoonma@163.com.}
\affiliation{
Zhengzhou Information Science and Technology Institute,zhengzhou,450002,China
}%


\date{\today}

\begin{abstract}
Computing the unit group and solving the principal ideal problem for
a number field are two of the main tasks in computational algebraic
number theory. This paper proposes efficient quantum algorithms for
these two problems when the number field has constant degree. We
improve these algorithms proposed by Hallgren by using a period
function which is not one-to-one on its fundamental period.
Furthermore, given access to a function which encodes the lattice, a
new method to compute the basis of an unknown real-valued lattice is
presented.
\end{abstract}

\maketitle

\section{Introduction}

Quantum algorithms can be used to realize a sub-exponential or even
exponential speed-up over known classical algorithms for some
mathematical problems by using Shor's\cite[Shor 1994]{Sho94}
algorithm framework. By extending the notion of period function,
Hallgren\cite[Hallgren 2002]{Hal02} showed how to approximate to the
period of an irrational periodic function. Moreover, Hallgren
applied the proposed technique to compute the regulator of a
real-quadratic field and solve the principal ideal problem in
polynomial time. Computing the regulator (Regulator Problem) and
solving the principal ideal problem (PIP) are interesting not only
from a pure mathematical point of view. Buchmann\cite[Buchmann
1990]{Buch90} and Williams proposed a Diffie-Hellman-like
cryptosystem whose security is based on PIP. Thus, if we could solve
the PIP, we will break the cryptosystem proposed by Buchmann. We
should choose a better cryptosystem if we assume that a large-scale
quantum computer can be build.

One small problem which arose during these computations was the
choice of the right approximation of natural logarithms. There was
no known way to choose the approximation in advance for a given
number field, so Schmidt\cite[Schmidt 2005]{Sch05} pointed out that
there remains a gap in Hallgren's\cite[Hallgren 2002]{Hal02}
algorithm for the quadratic case. Moreover, Schmidt closed a gap
left open by Hallgren and generalized Hallgren's work to
$\mathbb{Z}^r $. This generalized frame-work was then applied to
compute the unit group of an algebraic number field. Schmidt's
algorithm achieved an exponential speed-up over the best classical
deterministic algorithm. The problem was also independently solved
by Hallgren\cite[Hallgren 2005, Hallgren 2007]{Hal05, Hal07}
himself. Hallgren computed the unit group, solved the principal
ideal problem, and computed the class group, for constant degree
number fields, in polynomial time.

More recently, Schmidt\cite[Schmidt 2009]{Sch09} showed that the
regulator problem and the PIP in real-quadratic number fields can
also be solved by using functions which are always periodic but are
many-to-one on their fundamental period. They showed that Shor's
framework could compute the right period even in such a case with
constant success probability.

Inspired by Hallgren's original work, we show that the unit group
and the principal ideal problem for constant degree number fields,
can also be solved by using functions which are always periodic but
are many-to-one on their fundamental period lattice. In this paper,
we solve these problems for certain many-to-one functions whose
period are irrational and present more efficient algorithms for
these problems. The success probability for the unit group problem
is $(2^{7r + 1} r^{2r} )^{ - 1} $ from Schmidt\cite[Schmidt
2005]{Sch05} and $(2^{3r + 3} (r\log \Delta )^r )^{ - 1} $
 from Hallgren\cite[Hallgren 2005]{Hal05}, respectively. However, the probability from this paper
is at least $\left( {100 \cdot (3r)^{2r}  \cdot 5^r } \right)^{ - 1}
$, where $r$ is a constant and $\log \Delta  \geqslant r$.

The rest of this paper is organized as follows. In section 2, we
give a short overview of the quantum computation and the algebraic
number theory. In section 3, a quantum algorithm for computing the
unit group of a given number field will be presented. In section 4,
we propose an algorithm for the principal ideal problem. Conclusions
are given in section 5.

\section{Backgrounds}
\subsection{Quantum Computing}
First we give a brief introduction to quantum computation. Many
problems that have quantum algorithms with exponential speed-up over
the best known classical algorithm use the quantum Fourier transform
(QFT) as a subroutine. These problems can be reduced to the problem
of finding a basis of a period lattice $\Lambda $. We denote by
$\cdot  $ the dot product of two vectors and by the lattice
${\Lambda ^*}$ which is dual to $\Lambda $, i.e., $\Lambda ^{*}
=\left\{\mathbf v\in {\rm span}(\Lambda ) \left|\forall \mathbf u\in
\Lambda :\mathbf v\cdot  \mathbf u\in {\bf {\mathbb Z}}\right.
\right\}$. Generally speaking, if a basis of the dual lattice
$\Lambda ^{*} $ is known, one can compute a basis of the original
lattice $\Lambda $ by classical computer efficiently. So, it is
enough for the quantum algorithms to find an approximation of a
basis $\mathbf B$ for the dual lattice $\Lambda ^{*} $. Several
known quantum algorithms which achieved exponential speed-up are
based on this framework, such as Shor's factorization and discrete
logarithms algorithms, Hallgren's algorithms for pell's equation.

The framework for such an algorithm proceeds as follows: The quantum
computer uses two registers: one to store the input of the function
and the other to store the function value. Firstly, the quantum
computer creates a superposition of all possible states in the first
register, computes the function values and stores them in the second
register. Secondly, we measure the second register. By the laws of
quantum mechanics, the state of the quantum computer transforms into
$\sum _{\mathbf v\in L}{\left| \mathbf u+\mathbf v \right\rangle}
{\left| f(\mathbf u) \right\rangle}  $ where $\mathbf u$ is a random
vector and $L$ is a subset of $\Lambda $. Thirdly, the QFT and a
measurement are applied to the first register.  Now, we get a vector
from a basis of $\Lambda ^{*} $.

So, for a lattice $\Lambda $ with fixed dimension, we can get an
approximation of the basis ${\mathbf{B}}$ of the lattice $\Lambda
^{*} $ with fixed probability after running the subroutine above a
constant number of times. The QFT has an interesting and useful
property, known as shift invariance. i.e., the resulting
distribution is independent of which coset is started with. Thus,
the QFT always creates a superposition of value which approximates
the basis of $\Lambda ^{*} $ independent of $\mathbf u$.
Furthermore, after running the QFT to the register, the elements in
the superposition are almost uniformly distributed. More detail
about quantum computing, see Nielsen's\cite[Nielsen2000]{Niel00}
book.

\subsection{Algebraic number theory}
In this section we give the necessary background on algebraic number
theory. One can find almost all of the following facts from
Thiel's\cite[Thiel 1995]{Thiel95} work or Cohen's\cite[Cohen
1993]{Coh93} standard book on computational algebraic number theory.

A number field $K$ can be defined as a subfield of the complex
numbers $\mathbb{C}$ which is generated over the rational numbers
$\mathbb{Q}$ by an algebraic number, i.e.,  $K={\mathbb Q}(\theta )$
where $\theta $ where $\theta $ is a root of a monic irreducible
polynomial of degree $n$ with rational coefficients, which is called
the minimal polynomial of $\theta $. The number $n$ is called the
degree of $K$(over $\mathbb{Q}$). The signature of  $K$ is the pair
$(s,t) \in \left| \mathbb{Z} \right| \times \left| \mathbb{Z}
\right|$ , where $s$ is the number of real zeros of the minimal
polynomial of $\theta $ and $t$  is the number of pairs of nonreal
zeros; clearly, we have $s + 2t = n$. The signature is independent
of the choice of the generating polynomial and thus is an invariant
of the number field.

First we introduce some properties associated with number fields. In
the following, we shall always assume that $K={\mathbb Q}(\theta )$
is a number field of signature $(s,t)$. If $\theta _1 ,...,\theta
_n$ are the roots of the minimal polynomial of $\theta $, then there
are $n$ ways to embed the number field in $\mathbb{C}$. Let $m = s +
t$. An element in $K$ has $n$ conjugates, and $K$ has $m$ absolute
values, all of which correspond to the embeddings. Given any number
$\alpha  \in K$, $\alpha  = \sum\nolimits_{i = 0}^{n - 1} {a_i
\theta ^i } $ for some rational numbers $a_i  \in \mathbb{Q}$, let
$\alpha ^{(j)} $ denote the j-th conjugate of $\alpha $, i.e., the
image of $\alpha $ in the j-th embedding: $\alpha ^{(j)}  =
\sum\nolimits_{i = 0}^{n - 1} {a_i \theta _j^i } $. The j-th
absolute value $\left|  \cdot  \right|_j $ of a number  $\alpha $ is
a function of the absolute value in the j-th conjugate field:
$\left| \alpha  \right|_j  = \left\{ {\begin{array}{*{20}c}
   {\left| {\alpha ^{(j)} } \right|} & {1 \leqslant j \leqslant s}  \\
   {\left| {\alpha ^{(j)} } \right|^2 } & {s + 1 \leqslant j \leqslant m}  \\
\end{array} } \right.$,
where $\left| \alpha  \right|_j  = 0 \Leftrightarrow \alpha  = 0$.

An order ${\mathcal O}$ of a number field $K$ is a subring of
containing 1 that also is a module of $K$. Let ${\mathcal O}$  be an
order of a number field $K$. A number $\xi  \in {\mathcal O}$
 such that $\xi ^{ - 1}  \in {\mathcal O}$ is called a unit ${\mathcal O}$.
 The set of all units of ${\mathcal O}$ is a multiplicative abelian group that
 is called the unit group of ${\mathcal O}$ and is
denoted by  ${\mathcal O}^{*} $. By Dirichlet unit theorem, if we
set $r = s + t - 1$, we see that there exist $\varepsilon _1
,...\varepsilon _r $ such that every $\varepsilon  \in {\mathcal
O}^* $ can be written in a unique way as $\varepsilon  = \zeta
\varepsilon _1^{n_1 } ,...\varepsilon _r^{n_r } $, where $n_i  \in
\mathbb{Z}$ and $\zeta $ is a root of unity in $K$. So the unit
group in general will be isomorphic to $\mathbb{Z}^r $, together
with a root of unity. Given a number field of constant degree, the
root of unity can be computed efficiently by a classical computer.
So computing the unit group  ${\mathcal O}^{*} $ will mean computing
a fundamental system of units $\varepsilon _1 ,...\varepsilon _r $
that generate   ${\mathcal O}^{*} $.

\textbf{Definition 1} A fractional ${\mathcal O}$-ideal $I$ is a
non-zero free ${\bf {\mathbb Z}}$-submodule of $K$ such that there
exists a non-zero integer $d$ with $dI$ ideal of ${\mathcal O}$. An
ideal is said to be a principal ideal if there exists $x\in K$ such
that $I=x{\mathcal O}$.

\textbf{Definition 2} Let $I$ is a fractional ideal and $\alpha $ a
non-zero element of $I$. We will say that $\alpha $ is a minimum of
$I$ if, for all $\beta \in I$, we have $\forall i$, $\left|\beta
\right|_{i} <\left|\alpha \right|_{i} \Rightarrow \beta =0$, and the
set of all minima of ${\mathcal O}$ will be denote by ${\mathcal
M}_{{\mathcal O}} $. We will say that the ideal $I$ is reduced if
$l(I)$ is a minimum in $I$, where $I\cap {\mathbb
Q}=l(I){\bf{\mathbb Z}}$.

For a given ideal, there are an exponential number of minima in
general. A reduced ideal is important because it is possible to keep
the representation size bounded by a polynomial. The set of all
principal reduced ideals ${\mathcal R}_{{\mathcal O}} $ is precisely
the set of ideals $\frac{1}{\sigma } {\mathcal O}$ where $\sigma $
runs through all minima of ${\mathcal O}$.

\textbf{Definition 3} The logarithmic embedding of $K^{*} $ in
${\mathbb R}^{s+t} $ is the map ${\rm Log}$ which sends $\alpha $ to
\[{\rm Log:}\alpha \mapsto \left(\log \left|\alpha \right|_{1} ,...,\log \left|\alpha \right|_{s+t} \right).\]

\textbf{Definition 4 (Unit group problem)}. Given a number field
 $K$ and the ring of integers ${\mathcal O}$, find a system of
fundamental units of $K$.

\textbf{Lemma 1}\cite[Cohen 1993]{Coh93}\textbf{ }The image of the
group of units ${\mathcal O}^{*} $ under the logarithmic embedding
is a lattice(of rank \textit{r})in the hyperplane $\sum _{1\le i\le
s+t}\alpha _{i} =0$ of ${\mathbb R}^{r+1} $. The kernel of the
logarithmic embedding is exactly equal to the group of the roots of
unity in $K$.

Given the lattice $\Lambda $, one can get the group of unit
${\mathcal O}^{*} $ by classical computer efficiently. So it is
enough for us to find a basis of the lattice $\Lambda $.

\section{Computing the unit group}

\subsection{The periodic function}
By assigning to each point $\mathbf v$ in ${\mathbb Q}^{r} $ the
element of ${\mathcal R}_{{\mathcal O}} $ which is closest to
$\mathbf v$ mod $\Lambda $ we obtain a periodic function with period
lattice $\Lambda $. Unlike Hallgren's work, we consider many-to-one
periodic function, thus, stringent injectivity entirely discarded.

First we give the definition of the periodic function on ${\bf
{\mathbb Z}}^{r} $ hides $\Lambda $ for computing the unit group.
For some $N \in \mathbb{Z}$ we define $f_{N} $ as follows:
\[f_{N} :{\bf {\mathbb Z}}^{r} \to {\mathcal R}_{{\mathcal O}} :\mathbf v\mapsto I_{{\mathbf v\mathord{\left/ {\vphantom {\mathbf v N}} \right. \kern-\nulldelimiterspace} N} } =\frac{1}{\sigma ({\mathbf v\mathord{\left/ {\vphantom {\mathbf v N}} \right. \kern-\nulldelimiterspace} N} )} {\mathcal O}\]
Where $I_{{\mathbf v\mathord{\left/ {\vphantom {\mathbf v N}}
\right. \kern-\nulldelimiterspace} N} } =\frac{1}{\sigma ({\mathbf
v\mathord{\left/ {\vphantom {\mathbf v N}} \right.
\kern-\nulldelimiterspace} N} )} {\mathcal O}$ is the reduced ideal
such that $\sigma ({\mathbf v\mathord{\left/ {\vphantom {\mathbf v
N}} \right. \kern-\nulldelimiterspace} N} )$ is the minimum of
${\mathcal O}$ that minimizes $\left\| {\raise0.7ex\hbox{$ \mathbf v
$}\!\mathord{\left/ {\vphantom {v N}} \right.
\kern-\nulldelimiterspace}\!\lower0.7ex\hbox{$ N $}} -{\rm
Log}\sigma ({\mathbf v\mathord{\left/ {\vphantom {\mathbf v N}}
\right. \kern-\nulldelimiterspace} N} )\right\| _{2} $. Especially,
if there are two or more $\sigma ({\mathbf v\mathord{\left/
{\vphantom {\mathbf v N}} \right. \kern-\nulldelimiterspace} N} )$
meet the condition, we choose the right one by lexicographic
comparison.

The difference between function defined by Hallgren's and this paper
is that the injectivity in our function will be dropped entirely. By
the results of Hallgren\cite[Hallgren 2005]{Hal05} and
Schmidt\cite[Schmidt 2005]{Sch05}, and demonstrated in detail in
algorithm 6.2.20 \cite[Thiel 1995]{Thiel95}, one can compute the
reduced ideal that near the given point in polynomial time for
number fields with constant degree.

Next we will show that $f_N $ is periodic.

\textbf{Definition 5} Let $M \subset \mathbb{Z}^r $, the centre of
$M$ is one point ${\mathbf{p}} \in M$ such that for any
${\mathbf{p}}' \in M$, $\sum\nolimits_{{\mathbf{v}} \in M} {\left\|
{{\mathbf{p}} - {\mathbf{v}}} \right\|_2 }  \leqslant
\sum\nolimits_{{\mathbf{v}} \in M} {\left\| {{\mathbf{p}}' -
{\mathbf{v}}} \right\|_2 } $, especially, if there are two or more
${\mathbf{p}}$ meet the condition, we choose the right by one by
lexicographic.

\textbf{Lemma 2} Let $S_\sigma   = \{ {\mathbf{w}}' \in
\mathbb{Z}_q^r |f_N ({\mathbf{w}}') = \frac{1} {\sigma }{\mathcal
O}\} $ and ${\mathbf{v}} \in \mathbb{Z}_q^r $ belong to the same
fundamental parallelepiped of $N\Lambda $, ${\mathbf{w}}$
 is the centre of $S_\sigma  $. We denote the absolute of the discriminant of  ${\mathcal
O}$ by $\Delta _{\mathcal O} $
 . Then, for any ${\mathbf{n}} \in N\Lambda $, there exists
$\left\| {{\mathbf{\beta }}({\mathbf{w}},{\mathbf{n}})}
\right\|_\infty   \leqslant \frac{1} {4}\log \Delta _{\mathcal O} $
 such that the following is true,

(1) Let $\overline {\mathbf{v}}  = {\mathbf{v}} - {\mathbf{w}} =
(\overline v _1 ,\overline v _2 ,...,\overline v _r )$,
${\mathbf{\beta }}({\mathbf{w}},{\mathbf{n}}) = (\beta _1 ,\beta _2
,...\beta _r )$, for any $1 \leqslant i \leqslant r$, if $\left|
{\overline v } \right|_i  > \left| {\beta _i } \right|$, then
${\mathbf{w}} + \overline {\mathbf{v}}  + {\mathbf{n}} +
{\mathbf{\rho }}({\mathbf{w}},{\mathbf{n}}) \notin S_\sigma  $,
where $\left\| {{\mathbf{\rho }}({\mathbf{w}},{\mathbf{n}}))}
\right\|_\infty   \leqslant {1 \mathord{\left/
 {\vphantom {1 2}} \right.
 \kern-\nulldelimiterspace} 2}$.

(2) For ${\mathbf{n}},{\mathbf{n}}' \in N\Lambda $, $\max
_{{\mathbf{n}},{\mathbf{n}}'} \left\| {{\mathbf{\beta
}}({\mathbf{w}},{\mathbf{n}}) - {\mathbf{\beta
}}({\mathbf{w}},{\mathbf{n}}')} \right\|_\infty   \leqslant 2$.

Proof: (1) By lemma 5.1.14 proved in [9], the number $N$ of minima
in a box of side length $\frac{1} {4}\log \Delta _{\mathcal{O}} $
  satisfies
$1 \leqslant N \leqslant 4^n (\log \Delta _{\mathcal{O}} )^r $. So
the distance of two minimum is less than $\frac{1} {2}\log \Delta
_{\mathcal{O}} $, then, if $\left| {\overline v _i } \right| >
\frac{1} {4}\log \Delta _{\mathcal{O}} $, we have ${\mathbf{w}} +
\overline {\mathbf{v}}  + {\mathbf{n}} + {\mathbf{\rho
}}({\mathbf{w}},{\mathbf{n}}) \notin S_\sigma  $, i.e. $\left\|
{{\mathbf{\beta }}({\mathbf{w}},{\mathbf{n}})} \right\|_\infty
\leqslant \frac{1} {4}\log \Delta _{\mathcal{O}} $.

(2) Let ${\mathbf{n}} = N{\text{Log}}\varepsilon $ for some unit
$\varepsilon $. If $\sigma $ is the minimum closest to
$\frac{{{\mathbf{w}} + \overline {\mathbf{v}} }} {N}$, then in most
of case, $\varepsilon \sigma $ is the one closest to $\frac{{\left[
{{\mathbf{w}} + \overline {\mathbf{v}}  + {\mathbf{n}}} \right]}}
{N}$. Here $\left[  \cdot  \right]$ rounds to the closest integer
and is applied to the vector component-wise. If and only if
${\mathbf{w}} + \overline {\mathbf{v}} $ is in the boundary of
$S_\sigma  $, we can't determine whether ${\mathbf{w}} + \overline
{\mathbf{v}}  + {\mathbf{n}} + {\mathbf{\rho
}}({\mathbf{w}},{\mathbf{n}}) \in S_\sigma  $ holds. Then due to
rounding, for different ${\mathbf{n}},{\mathbf{n}}' \in N\Lambda $,
$\max _{{\mathbf{n}},{\mathbf{n}}'} \left\| {{\mathbf{\beta
}}({\mathbf{w}},{\mathbf{n}}) - {\mathbf{\beta
}}({\mathbf{w}},{\mathbf{n}}')} \right\|_\infty   \leqslant 2$. ¡õ

\subsection{The algorithm}
In this section we present a method to compute a basis for a
constant dimensional lattice hidden by a function, and to solve some
instances of the hidden subgroup problem over ${\mathbb R}^{r} $.

Given a function hiding a lattice $\Lambda $ we will show how to
compute a basis for the dual lattice $\Lambda ^{*} $. To compute a
basis for $\Lambda $ we need the lattice be well conditioned. A
lattice is well conditioned if a matrix $\mathbf B$ whose columns
form a basis for $\Lambda $ is well conditioned, i.e., if $\left\|
\mathbf B\right\| \cdot \left\| \mathbf B^{-1} \right\| $ is
bounded.

We denote the discriminant of $K$ by $\Delta $. For the purposes of
analyzing running times, it is customary to use $\Delta $ as input,
and an algorithm is polynomial or exponential if it is in $O((\log
\Delta )^{c} )$ or $O(\Delta ^{c'} )$ for some $c,c'\in {\mathbb
R}$, respectively, where the $O$-constants might depend
exponentially on $n$.

Next we propose an algorithm to find an $\varepsilon $-approximation
to a basis of $\Lambda ^{*} $.

Let $N\gg (\log \Delta )^{r} $ and $q\gg \det (N\Lambda )$ be a
power of 2. Now we present our algorithm. The complete analysis will
be given later.

------------------------------------------------------------------

\textbf{Algorithm 1}

------------------------------------------------------------------

Input: Number field $K$ and the ring of integers ${\mathcal O}$

Out: A set of vectors approximating a basis for $\Lambda ={\rm
Log}{\mathcal O}^{*} $

1)(Create superposition)

$ \to \frac{1} {{\sqrt {q^r } }}\sum\limits_{w_1  = 0}^{q - 1} {...}
\sum\limits_{w_r  = 0}^{q - 1} {\left| {w_1 } \right\rangle
...\left| {w_r } \right\rangle \left| 0 \right\rangle } $;

2) (Compute function )

$ \to \frac{1} {{\sqrt {q^r } }}\sum\limits_{w_1  = 0}^{q - 1} {...}
\sum\limits_{w_r  = 0}^{q - 1} {\left| {w_1 } \right\rangle
...\left| {w_r } \right\rangle \left| {f_N ({\mathbf{w}})}
\right\rangle } $; where ${\mathbf{w}} = (w_1 ,...w_r )$.

3) (Measure the second register)

$ \to \frac{1} {{\sqrt T }}\sum\limits_{{\mathbf{n}} \in L}
{\sum\limits_{i = 1}^{vol({\mathbf{\beta
}}({\mathbf{w}},{\mathbf{n}}))} {\left| {{\mathbf{w}} + \overline
{\mathbf{v}} _i  + {\mathbf{n}} + {\mathbf{\rho
}}({\mathbf{w}},{\mathbf{n}})} \right\rangle \left| {f_N
({\mathbf{w}})} \right\rangle } } $

With a random ${\mathbf{w}}$, $T = {\text{card}}\left\{
{{\mathbf{w}}' \in \mathbb{Z}_q^r |} \right.f_N ({\mathbf{w}}') =
\left. {f_N ({\mathbf{w}})} \right\}$, $vol({\mathbf{\beta
}}({\mathbf{w}},{\mathbf{n}}))$ is the number of ${\mathbf{v}}_i  =
(v_{i1} ,...,v_{ir} ) \in \mathbb{Z}_q^r $ such that $\left|
{\overline v _{ij} } \right| = \left| {v_{ij}  - w_j } \right| <
\beta _j $ and $f_N ({\mathbf{w}} + \overline {\mathbf{v}} _i  +
{\mathbf{n}} + {\mathbf{\rho }}({\mathbf{w}},{\mathbf{n}})) = f_N
({\mathbf{w}})$; $L = \left\{ {{\mathbf{n}} \in N\Lambda
|{\mathbf{w}} + \overline {\mathbf{v}} _i  + {\mathbf{n}} +
{\mathbf{\rho }}({\mathbf{w}},{\mathbf{n}}) \in \mathbb{Z}_q^r }
\right\}$

Test whether $f({\mathbf{w}})$ lie in the set for which periodicity
can be guaranteed, if not, restart;

4) (Apply the QFT to the first register)

$\begin{gathered}
   \to \frac{1}
{{\sqrt {(kq)^r T} }}\sum\nolimits_{{\mathbf{c}} \in
\mathbb{Z}_{qk}^r } {\sum\limits_{{\mathbf{n}} \in L}
{\sum\limits_{i = 1}^{vol({\mathbf{\beta
}}({\mathbf{w}},{\mathbf{n}}))} {\exp \left( {\frac{{2\pi i}}
{{qk}}} \right.\left( {{\mathbf{w}} + \overline {\mathbf{v}} _i } \right.} } }  \hfill \\
   + {\mathbf{n}}\left. {\left. { + {\mathbf{\rho }}({\mathbf{w}},{\mathbf{n}})} \right)
   \cdot {\mathbf{c}}} \right)\left| {\mathbf{c}} \right\rangle \left| {f_N ({\mathbf{w}})} \right\rangle  \hfill \\
\end{gathered} $

Where $k$ is a constant that will be determined later;

5) Measure and return the first register ${\mathbf{c}}$;

6) Repeat the procedure; compute a basis of $(N\Lambda )^* $
 from the spanning set of vectors;

7) Compute a basis for $\Lambda $ classically.

------------------------------------------------------------------

\textbf{Notes}: We will explain the constant $k$  appearing in step
4. In algorithm 1, just run the QFT over $\mathbb{Z}_q^r $ as usual
does not appear to be enough to recover the dual lattice. To
overcome this problem we use constant  $k$ to run the QFT, i.e. we
'zero-fill', to compute the larger domain $\mathbb{Z}_{qk}^r $, with
the additional part of the domain taking zero values. This
constraint also helps us to confine the errors caused by the factor
${\mathbf{\rho }}({\mathbf{w}},{\mathbf{n}})$ in the function $f_N
$. This type of operation has been studied by Hallgren\cite[Hallgren
2005, Hales 1999]{Hal05,Hale99}.

Algorithm 1 is a typical algorithm for hidden subgroup problem.
After apply the QFT and measure the first register, we can get an
appropriate ${\mathbf{c}}$. Thus, one vector from a basis of
$(N\Lambda )^* $ can be efficiently obtained.

Next, we will present the complete analysis for success probability.

We want to estimate the probability to measure ${\mathbf{c}}$  with
$\left\| {{\raise0.7ex\hbox{${\mathbf{c}}$} \!\mathord{\left/
 {\vphantom {{\mathbf{c}} {qk}}}\right.\kern-\nulldelimiterspace}
\!\lower0.7ex\hbox{${qk}$}} - {\mathbf{n}}^* } \right\|_\infty
\leqslant \frac{1} {{2qk}}$. To keep the influence of disturbing
${\mathbf{\rho }}({\mathbf{w}},{\mathbf{n}})$ small, we consider
only "small"  ${\mathbf{c}}$ and restart the algorithm if
${\mathbf{c}}$ is too big. For simplify analysis, without loss of
generality, let ${\mathbf{\beta }}({\mathbf{w}},{\mathbf{n}}) =
(\beta _1 ,\beta _2 ,...\beta _r )$ and $\beta _i  = \beta $, $(1
\leqslant i \leqslant r)$, i.e., $S_\sigma  $ be a multidimensional
sphere and $\beta $ is radius.

\textbf{Lemma 3} Let  $k = 3r$, ${\raise0.7ex\hbox{${\mathbf{c}}$}
\!\mathord{\left/
 {\vphantom {{\mathbf{c}} {qk}}}\right.\kern-\nulldelimiterspace}
\!\lower0.7ex\hbox{${qk}$}} = {\mathbf{n}}^*  + {\mathbf{\delta
}}({\mathbf{c}})$,

${\mathbf{C}}{\text{ = }}\left\{ {{\mathbf{c}} \in \mathbb{Z}_{qk}^r
|\left\| {\mathbf{c}} \right\|_\infty   < \frac{q} {{5 \cdot (\beta
+ 1)}},{\raise0.7ex\hbox{${\mathbf{c}}$} \!\mathord{\left/
 {\vphantom {{\mathbf{c}} {qk}}}\right.\kern-\nulldelimiterspace}
\!\lower0.7ex\hbox{${qk}$}} - {\mathbf{\delta }}({\mathbf{c}}) \in
(N\Lambda )^* } \right\}$, where $\left\| {{\mathbf{\delta
}}({\mathbf{c}})} \right\|_\infty \leqslant \frac{1} {{2qk}}$, then
the probability to get a vector from a basis of $(N\Lambda )^* $
 is at least
$\left( {100 \cdot (3r)^{2r}  \cdot 5^r } \right)^{ - 1} $.

Proof. The QFT is shift invariant.

So for probability estimation we can assume ${\mathbf{w}} =
{\mathbf{0}}$. The probability to obtain a ${\mathbf{c}} \in
{\mathbf{C}}$ is $\frac{1} {{(kq)^r T}}\left|
{\sum\limits_{{\mathbf{n}} \in L} {\sum\limits_{i = 1}^{vol(\beta
({\mathbf{w}},{\mathbf{n}}))} {\exp \left( {\frac{{2\pi i}}
{{qk}}\left( {{\mathbf{w}} + \overline {\mathbf{v}} _i  +
{\mathbf{n}} + {\mathbf{\rho }}({\mathbf{w}},{\mathbf{n}})} \right)
\cdot {\mathbf{c}}} \right)} } } \right|$ =
\begin{equation}
\frac{1} {{(kq)^r T}}\left| {\sum\limits_{{\mathbf{n}} \in L}
{\sum\limits_{i = 1}^{vol(\beta ({\mathbf{w}},{\mathbf{n}}))} {\exp
\left( {\frac{{2\pi i}} {{qk}}\left( {\overline {\mathbf{v}} _i  +
{\mathbf{n}} + {\mathbf{\rho }}({\mathbf{w}},{\mathbf{n}})} \right)
\cdot {\mathbf{c}}} \right)} } } \right|^2
\end{equation}
let

$\begin{gathered}
  s = \left( {\overline {\mathbf{v}} _i  + {\mathbf{n}} + {\mathbf{\rho }}({\mathbf{w}},{\mathbf{n}})} \right)
  \cdot {\raise0.7ex\hbox{${\mathbf{c}}$} \!\mathord{\left/
 {\vphantom {{\mathbf{c}} {kq}}}\right.\kern-\nulldelimiterspace}
\!\lower0.7ex\hbox{${kq}$}} \hfill \\
   = \overline {\mathbf{v}} _i  \cdot {\raise0.7ex\hbox{${\mathbf{c}}$} \!\mathord{\left/
 {\vphantom {{\mathbf{c}} {kq}}}\right.\kern-\nulldelimiterspace}
\!\lower0.7ex\hbox{${kq}$}} + {\mathbf{n}} \cdot ({\mathbf{n}}^*  +
{\mathbf{\delta }}({\mathbf{c}})) + {\mathbf{\rho
}}({\mathbf{w}},{\mathbf{n}}) \cdot
{\raise0.7ex\hbox{${\mathbf{c}}$} \!\mathord{\left/
 {\vphantom {{\mathbf{c}} {kq}}}\right.\kern-\nulldelimiterspace}
\!\lower0.7ex\hbox{${kq}$}} \hfill \\
   = \overline {\mathbf{v}} _i  \cdot {\raise0.7ex\hbox{${\mathbf{c}}$} \!\mathord{\left/
 {\vphantom {{\mathbf{c}} {kq}}}\right.\kern-\nulldelimiterspace}
\!\lower0.7ex\hbox{${kq}$}} + {\mathbf{n}} \cdot {\mathbf{n}}^*  +
{\mathbf{n}} \cdot {\mathbf{\delta }}({\mathbf{c}}) + {\mathbf{\rho
}}({\mathbf{w}},{\mathbf{n}}) \cdot
{\raise0.7ex\hbox{${\mathbf{c}}$} \!\mathord{\left/
 {\vphantom {{\mathbf{c}} {kq}}}\right.\kern-\nulldelimiterspace}
\!\lower0.7ex\hbox{${kq}$}} \hfill \\
\end{gathered} $

Since $\left\| {\mathbf{n}} \right\|_\infty   < q$, $\left\|
{\mathbf{c}} \right\|_\infty   < \frac{q} {{5 \cdot (\beta + 1)}}$,
$\left\| {{\mathbf{\delta }}({\mathbf{c}})} \right\|_\infty
\leqslant \frac{1} {{2qk}}$ , we have $s\bmod 1$
 =
$\overline {\mathbf{v}} _i  \cdot {\raise0.7ex\hbox{${\mathbf{c}}$}
\!\mathord{\left/
 {\vphantom {{\mathbf{c}} {kq}}}\right.\kern-\nulldelimiterspace}
\!\lower0.7ex\hbox{${kq}$}} + {\mathbf{n}} \cdot {\mathbf{\delta
}}({\mathbf{c}}) + {\mathbf{\rho }}({\mathbf{w}},{\mathbf{n}}) \cdot
{\raise0.7ex\hbox{${\mathbf{c}}$} \!\mathord{\left/
 {\vphantom {{\mathbf{c}} {kq}}}\right.\kern-\nulldelimiterspace}
\!\lower0.7ex\hbox{${kq}$}}$

$ \leqslant r\frac{{\left\| {\overline {\mathbf{v}} _i }
\right\|_\infty   \cdot q}} {{5 \cdot (\beta  + 1) \cdot kq}} +
r\frac{q} {{2qk}} + r\frac{q} {{10 \cdot (\beta  + 1) \cdot kq}}$

$ \leqslant \frac{r} {{5k}} + \frac{r} {{2k}} + \frac{r} {{10k \cdot
(\beta  + 1)}}$

From the definition of ${\mathbf{\beta
}}({\mathbf{w}},{\mathbf{n}})$, we know that $2k(\beta  + 1) \gg
2k$. So if $k = 3r$, then $s\bmod 1 \leqslant \frac{7} {{30}} +
\frac{1} {{30 \cdot (\beta  + 1)}} \approx \frac{7} {{30}}$. It
follows that the angle between the vectors $\exp \left( {\frac{{2\pi
i}} {{qk}}\left( {\overline {\mathbf{v}} _i  + {\mathbf{n}} +
{\mathbf{\rho }}({\mathbf{w}},{\mathbf{n}})} \right) \cdot
{\mathbf{c}}} \right)$ in Eq.(1) is  $[ - \frac{7} {{15}}\pi
,\frac{7} {{15}}\pi ]$. So the absolute value of the sum is larger
than $\frac{T} {{(3rq)^r }}\left| {\cos \frac{7} {{15}}\pi }
\right|^2 \approx \frac{T} {{100 \cdot 3^r r^r q^r }}$; Furthermore,
applying Proposition 8.7 in \cite[Micciancio 2002]{Mic02}, we have
that ${\text{card}}\left\{ {{\mathbf{n}} \in L} \right\} \approx
\frac{{q^r }} {{\det (N\Lambda )}}$, so $T = {\text{card}}\left\{
{{\mathbf{w}}' \in \mathbb{Z}_q^r |f_N ({\mathbf{w}}') = f_N
({\mathbf{w}})} \right\}$

$\begin{gathered}
   = \sum\limits_{i = 1}^{vol(\beta ({\mathbf{w}},{\mathbf{n}}))} {{\text{card}}\left\{ {{\mathbf{n}} \in L} \right\}}  \hfill \\
   \geqslant vol(\beta  - 1) \cdot \frac{{q^r }}
{{\det (N\Lambda )}} \hfill \\
\end{gathered} $

Next we approximate the cardinality of ${\mathbf{C}}$, We have

$\begin{gathered}
  {\text{card}}{\mathbf{C}} \geqslant {\text{card}}\left\{ {{\mathbf{c}} \in \mathbb{Z}_{qk}^r |\left\| {\mathbf{c}} \right\|_\infty   < \frac{q}
{{5 \cdot (\beta  + 1)}},{\raise0.7ex\hbox{${\mathbf{c}}$}
\!\mathord{\left/
 {\vphantom {{\mathbf{c}} {qk}}}\right.\kern-\nulldelimiterspace}
\!\lower0.7ex\hbox{${qk}$}} - {\mathbf{\delta }}({\mathbf{c}})} \right. \hfill \\
  \left. { \in (N\Lambda )^* } \right\} \approx \frac{{\det (N\Lambda )}}
{{(3r \cdot 5(\beta  + 1))^r }} \hfill \\
\end{gathered} $;
$\frac{{vol(\beta  - 1)}} {{(\beta  + 1)^r }} \approx 1$. Thus, the
probability $P$ to measure a 'good' ${\mathbf{c}}$ is larger than
$\left( {100 \cdot (3r)^{2r}  \cdot 5^r } \right)^{ - 1} $. So we
can obtain a vector from a basis of $(N\Lambda )^* $ from
${\mathbf{c}}$ . ¡õ

From lemma 2 in\cite[Schmidt 2005]{Sch05}, we need only a polynomial
repetition of algorithm 1 to get a basis for $(N\Lambda )^* $.

\textbf{Lemma 4}\cite[Schmidt 2005]{Sch05} Let $\Lambda $  be a
lattice of a fixed rank $r$. Then for $B_1  \in \mathbb{R},B_1  >
10\sqrt r \lambda _r (\Lambda )$, there is an algorithm which does
the following $O({\text{poly}}\log (\det (\Lambda )))$. It samples
at most random vectors $\lambda $
 from
$\Lambda  \cap \left\{ {{\mathbf{x}} \in \mathbb{R}^r |0 \leqslant
x_i  < B_1 ,i = 0,...,r} \right\}$
  and
outputs with probability exponentially close to one a set of vectors
from $\Lambda $
 which generate  $\Lambda $.

\textbf{Theorem 1} Algorithm 1 computes the unit group
${\mathcal{O}}^* $  of a constant degree number field $K$  in
quantum polynomial time.

Proof. The probability only depends on the degree of the number
fields by lemma 3. So, keep the degree fixed, we need only a
polynomial repetition of the above algorithm to get a generating set
for $(N\Lambda )^* $, the polynomial time bound is clear from lemma
4.      ¡õ

\section{The Principal Ideal Problem}
\textbf{Definition 6 (Principal ideal problem)} Given an ideal $I$
 of ${\mathcal{O}}$, determine whether or not it is a principal ideal, and if it is,
compute $\alpha  \in K$ such that $I = \alpha {\mathcal{O}}$.

Given a reduced principal ideal $I=\alpha {\mathcal O}=I_{\theta }
$, where $\theta ={\rm Log}\alpha $, define the function

$g_{N} :{\bf {\mathbb Z}}\times {\bf {\mathbb Z}}^{r} \to {\mathcal
R}_{{\mathcal O}} $ by $g_{N} (a,\mathbf v)=I_{a\theta -{\mathbf
v\mathord{\left/ {\vphantom {\mathbf v N}} \right.
\kern-\nulldelimiterspace} N} } $. The ideal $I_{a\theta -{\mathbf
v\mathord{\left/ {\vphantom {\mathbf v N}} \right.
\kern-\nulldelimiterspace} N} } $ can be computed efficiently by
multiplying $I^{a} $ and $I_{-{\mathbf v\mathord{\left/ {\vphantom
{\mathbf v N}} \right. \kern-\nulldelimiterspace} N} } $.
Furthermore, the function $g_{N} $ has period lattice
$\overline{\Lambda }$.

Where $\overline{\Lambda }=\left\{(b,\eta )\subseteq {\bf {\mathbb
Z}}\times {\mathbb R}^{r} |b\theta - {\eta  \mathord{\left/
 {\vphantom {\eta  N}} \right. \kern-\nulldelimiterspace} N} \in \Lambda \right\}$ and
one of its basis is $(1,N\theta ),(0,\mathbf v_{1} ),...,(0,\mathbf
v_{r} )$. Here $\mathbf v_{i} $ $(1 \leqslant i \leqslant r)$
 are one basis of the lattice
$N\Lambda $. Let ${\mathbf{e}} = (a,{\mathbf{v}})$ is a $r + 1$
dimensional vector, then we can denote ${g_N}(a,{\mathbf{v}})$ by
${g_N}({\mathbf{e}})$. Similarly, we give an algorithm to solve the
principal ideal problem.

------------------------------------------------------------------

\textbf{Algorithm 2}

------------------------------------------------------------------

Input: Number field $K$, the ring of integers ${\mathcal O}$ and a
reduced ideal $I$

Output: ${\text{Log}}\alpha $ if $I$ is a principal ideal, i.e. $I =
\alpha {\mathcal{O}}$; else `not principal'

 1)Create superstition and compute function
$g_N ({\mathbf{e}})$,

$ \to \frac{1} {{\sqrt {q^{r + 1} } }}\sum\limits_{{\mathbf{e}} \in
\mathbb{Z}_q^{r + 1} } {\left| {\mathbf{e}} \right\rangle \left|
{g_N ({\mathbf{e}})} \right\rangle } $

where ${\mathbf{e}} = (e_1 ,e_2 ,...e_{r + 1} )$

2) Measure the second register

$ \to \frac{1} {{\sqrt S }}\sum\limits_{{\mathbf{m}} \in
{\mathbf{M}}} {\sum\limits_{i = 1}^{vol({\mathbf{\beta
}}'({\mathbf{e}},{\mathbf{m}}))} {\left| {{\mathbf{e}} + \overline
{\mathbf{f}} _i  + {\mathbf{m}} + {\mathbf{\omega
}}({\mathbf{e}},{\mathbf{m}})} \right\rangle \left| {g_N
({\mathbf{e}})} \right\rangle } } $,

With a random ${\mathbf{e}} \in \mathbb{Z}_q^{r + 1} $, $S =
{\text{card}}\{ {\mathbf{e}}' \in \mathbb{Z}_q^{r + 1} |g_N
({\mathbf{e}}') = g_N ({\mathbf{e}})\} $, $vol({\mathbf{\beta
}}'({\mathbf{e}},{\mathbf{m}}))$ is the number of   such that
$\left| {\overline f _{ij} } \right| = \left| {f_{ij}  - e_j }
\right| < \beta _j '$ and $g_N ({\mathbf{e}} + \overline
{\mathbf{f}} _i  + {\mathbf{m}} + {\mathbf{\omega
}}({\mathbf{e}},{\mathbf{m}})) = g_N ({\mathbf{e}})$; ${\mathbf{M}}
= \{ {\mathbf{m}} \in \overline \Lambda  |{\mathbf{e}} + \overline
{\mathbf{f}} _i  + {\mathbf{m}} + \omega ({\mathbf{e}},{\mathbf{m}})
\in \mathbb{Z}_q^{r + 1} \} $.

Test whether $g_N ({\mathbf{e}})$  lie in the set for which
periodicity can be guaranteed, if not, restart;

3) Apply the QFT to the first register

$\begin{gathered}
  \frac{1}
{{\sqrt {(kq)^{r + 1} S} }}\sum\nolimits_{{\mathbf{c}} \in
\mathbb{Z}_{qk}^{r + 1} } {\sum\nolimits_{{\mathbf{m}} \in
{\mathbf{M}}} {\sum\nolimits_{i = 1}^{vol({\mathbf{\beta
}}'({\mathbf{e}},{\mathbf{m}}))} {\exp \left( {\frac{{2\pi i}}
{{qk}}} \right.} } }  \hfill \\
  \left. {\left( {{\mathbf{e}} + \overline {\mathbf{f}} _i  +
  {\mathbf{m}} + \omega ({\mathbf{e}},{\mathbf{m}})} \right)
  \cdot {\mathbf{c}}} \right)\left| {\mathbf{c}} \right\rangle \left| {g_N ({\mathbf{e}})} \right\rangle  \hfill \\
\end{gathered} $

4) Measure the first register, return ${\mathbf{c}}$;

5) Repeat the procedure, compute a basis of $\overline \Lambda
$£¬pick any two of them, ${\mathbf{c}} = (c,{\mathbf{f}}_1 )$,
${\mathbf{d}} = (d,{\mathbf{f}}_2 )$ such that $\gcd (c,d) = 1$;

6) Euclidean algorithm compute the linear combination make the first
coordinate equal 1,then we have $(1,{\mathbf{u}}) \in \bar \Lambda
$, therefore ${\mathbf{u}} = N{\text{Log}}\varepsilon \alpha $
 for some $\varepsilon $, where
$I = \varepsilon \alpha {\mathcal{O}}$;

7) Reduce $\mathbf u$ modulo the basis of $N\Lambda $, give an
optional ${\mathbf{\theta }}'$, if ${\mathbf{\theta }}'$ is an
approximation of ${\mathbf{\theta }}$, return it, else return `not
principal';

------------------------------------------------------------------

\textbf{Theorem 2 }Algorithm 2 works correctly as specified and
succeeds with constant probability. The principal ideal problem for
a constant number field can be solved in polynomial time by running
Algorithm 2.

Proof. Algorithm 2 compute a basis of $\overline \Lambda  $
 is obvious. There is not a unique generator, since $\varepsilon I=I$ for any unit
$\varepsilon \in {\mathcal O}^{*} $. Given any ideal a candidate
generator $\alpha' $ can be computed by running the algorithm. Then
we can compute $\alpha' {\mathcal O}$ by classical computers
efficiently. The result is $I$ if and only if $I$ is principal.
Furthermore, from the prime number theorem, the probability to
obtain two different non-zero vectors with the first coordinate
coprime is at least ${1 \mathord{\left/ {\vphantom {1 {\ln
q}}}\right.
 \kern-\nulldelimiterspace} {\ln q}}$. So we can obtain a correct result with pre-determined probability.

\section{Conclusions}
 In this paper, we solve two problems in computational
algebraic number theory. We have proposed algorithms to compute the
period lattice of many-to-one periodic functions, and applied the
technique to the computation of the unit group of a finite extension
$K$ of $\mathbb{Q}$. Furthermore, we extend the algorithm to solve
the PIP. The algorithm prints a correct result with pre-determined
probability. Its success probability can be arbitrarily increased by
repeating the algorithm. Thus the algorithm can be applied to attack
crypto-systems that rely on the difficulty of the principal ideal
problem yielding a better idea about which parameter sizes for these
crypto-systems remain secure in the presence of quantum
computers.This is due to the facts that the function value is a
reduced ideal and not a pair of an ideal and a distance.

Here we will discuss a few more open problems. The main problem is
that we haven't attempted to minimize the influence of the degree of
number field on the run-time which is unavoidably exponential now.
It is still an open problem whether or not there exist quantum
algorithms that solve these problems for arbitrary degree number
field. The other problems will be computing the class group for a
given number field by many-to-one function. Furthermore, finding
another practical problem which realize a exponential speed-up by
the proposed technique is more challengingly.

\end{document}